\documentclass[twocolumn,showpacs,superscriptaddress,showpacs,floatfix]{revtex4}

\usepackage{epsfig,color}
\usepackage{graphicx}
\usepackage{dcolumn}
\usepackage{bm}
\usepackage{setspace}

\begin{document}

\title{Degenerate versus semi-degenerate transport in a correlated 2D hole system}

\author{Richard L.J. Qiu}
\author{Xuan P. A. Gao}
\email{xuan.gao@case.edu}
 \affiliation{Department of Physics, Case Western Reserve University, Cleveland, OH 44106}
\author{Loren N. Pfeiffer}
\author{Ken W. West}
\affiliation{Department of Electrical Engineering, Princeton
University, Princeton, NJ 08544}

\begin{abstract}
It has been puzzling that the resistivity of high mobility
two-dimensional(2D) carrier systems in semiconductors with low
carrier density often exhibits a large increase followed by a
decrease when the temperature ($T$) is raised above a
characteristic temperature comparable with the Fermi temperature
($T_F$). We find that the metallic 2D hole system (2DHS) in GaAs
quantum well (QW) has a linear density ($p$) dependent
conductivity, $\sigma\approx e\mu^*(p-p_0)$, in both the
degenerate ($T\ll T_F$) and semi-degenerate ($T\sim T_F$) regimes.
The $T$-dependence of $\sigma(p)$ suggests that the metallic
conduction (d$\sigma$/d$T<$0) at low $T$ is associated with the
increase in $\mu^*$, the effective mobility of itinerant carriers.
However, the resistivity decrease in the semi-degenerate regime
($T>T_F$) is originated from the reduced $p_0$, the density of
immobile carriers in a two-phase picture.

\end{abstract}
\date{\today}
 \maketitle

The electron transport in 2D electron systems has been a research
focus for long time\cite{Lee&rama}. A pioneering work on the
one-parameter scaling theory of localization \cite{scaling79}
concluded that all non-interacting disordered 2D electronic
systems have to be localized in zero magnetic field ($B$=0). The
application of the celebrated scaling theory of localization and
the Fermi liquid (FL) model in strongly interacting 2D systems,
however, was questioned by a number of experimental observations
of an apparent metallic state and metal-insulator transition (MIT)
in various 2D electron or hole systems with low density and high
mobility \cite{mitreview}. Although low carrier density implies a
large value of $r_s$ (the ratio between Coulomb potential energy
and kinetic energy) or strong correlation effects, different
opinions exist on how the strong correlations affect the carrier
transport in dilute 2D systems and what mechanism causes the
metallic transport
\cite{philips,hexie,wc,vlad,punnoose,spivakkivelson,altshuler,zna,dassarma}.

After extensive experimental studies of transport in various
dilute 2D carrier systems in semiconductors, one salient feature
stands out in the 2D metallic transport phenomena. For densities
above the critical density $p_c$, the temperature dependent
resistivity $\rho(T)$ is often non-monotonic: $\rho$ first
increases and then decreases as the temperature is raised above a
characteristic temperature $T^*\sim T_F$. Such non-monotonic
behavior in $\rho(T)$ when low density 2D system becomes
semi-degenerate has been observed in all the three most widely
studied systems: n-Si \cite{skvprb99,Lai}, p-GaAs
\cite{Haneinprl,millsprl,gaoprl} and n-GaAs \cite{Lilly}. This
sign change in d$\rho$/d$T$ at $T^*$ is generic for the 2D
metallic state if the phonon scattering contribution to
resistivity does not overwhelm the impurity-scattering induced
$\rho$ in the semi-degenerate regime \cite{millsprl,gaophonon}.
The existence of a non-monotonic $\rho(T)$ is essential in many
leading theoretical explanations for the 2D metallic state
\cite{vlad,punnoose,spivakkivelson,dassarma}. Therefore, to
further distinguish the mechanisms of the 2D metallic state, it
would be desirable to address experimentally the transport and
scattering processes as the system crosses over from degenerate
($T\ll T_F$) to semi-degenerate ($T\sim T_F$) regime. In addition,
transport of 2D electron fluids with $r_s\gg$1 in the
semi-degenerate regime is interesting in its own right. In this
seldom studied regime, non-Boltzmann type transport like
hydrodynamics may play an important role
\cite{spivakkivelson,novikov}.

Here we compare the density dependence of conductivity in the
degenerate and semi-degenerate regimes for a low density 2DHS with
strong interactions ($r_s>$18 for the densities covered in this
experiment \cite{mass}). In the metallic state, our 2DHS in 10nm
wide GaAs QWs exhibits a pronounced non-monotonic $\rho(T)$
associated with the degenerate to semi-degenerate crossover and a
strong low $T$ metallicity\cite{gaoprl,gao06}, due to the stronger
confinement and smaller phonon scattering contribution to the
resistivity in narrow QWs\cite{millsprl,gaoprl,gaophonon}. The
particular focus of this paper is on understanding the
non-monotonic $\rho(T)$ of correlated 2DHS from the temperature
dependence of $\sigma(p)$ in the {\it high conductivity regime}
($\sigma\gg e^2/h$). In such metallic regime away from the
critical point of MIT, we find that the conductivity has a
Drude-like linear density dependence, $\sigma(p)\approx
e\mu^*(p-p_0)$, consistent with a two-phase mixture picture where
the total conductivity is dominated by mobile carriers with
mobility $\mu^*$ and density $p-p_0$. The $\sigma(p)$ data at
different $T$ further suggest that the resistivity change on the
two sides of the non-monotonic $\rho(T)$ of low density 2D systems
have distinct origins: one comes from $\mu^*(T)$ and the other is
a result of $p_0(T)$. The low $T$ metallic conduction in the
degenerate regime is accompanied by a sharply increasing
$\mu^*(T)$ as $T$ decreases. On the other hand, the resistivity
change of 2DHS in the non-degenerate regime is dominated by a
$p_0$ that decreases rapidly as $T$ increases.

Transport measurements were performed on a 2DHS in two 10nm wide
GaAs QW samples similar to the ones used in our previous
studies\cite{gaophonon,gao06,gaoHall}. The samples were grown on
(311)A GaAs wafer using Al$_{.1}$Ga$_{.9}$As barriers and
symmetrically placed Si delta-doping layers. The metal backgate
used to tune the hole density was about 0.15mm away from the QW,
such that the Coulomb interaction between holes are unscreened by
the gate and remains long-range. The samples were prepared in the
form of a Hall bar, of approximate dimensions 2$\times$9mm$^{2}$,
with diffused In(1\%Zn) contacts. The measurement current was
applied along the high mobility [\={2}33] direction and kept low
such that the power delivered on the sample was less than
3fWatts/cm$^2$ to avoid over heating the holes \cite{gaophonon}.
\begin{figure}[btph]
\centerline{\psfig{file=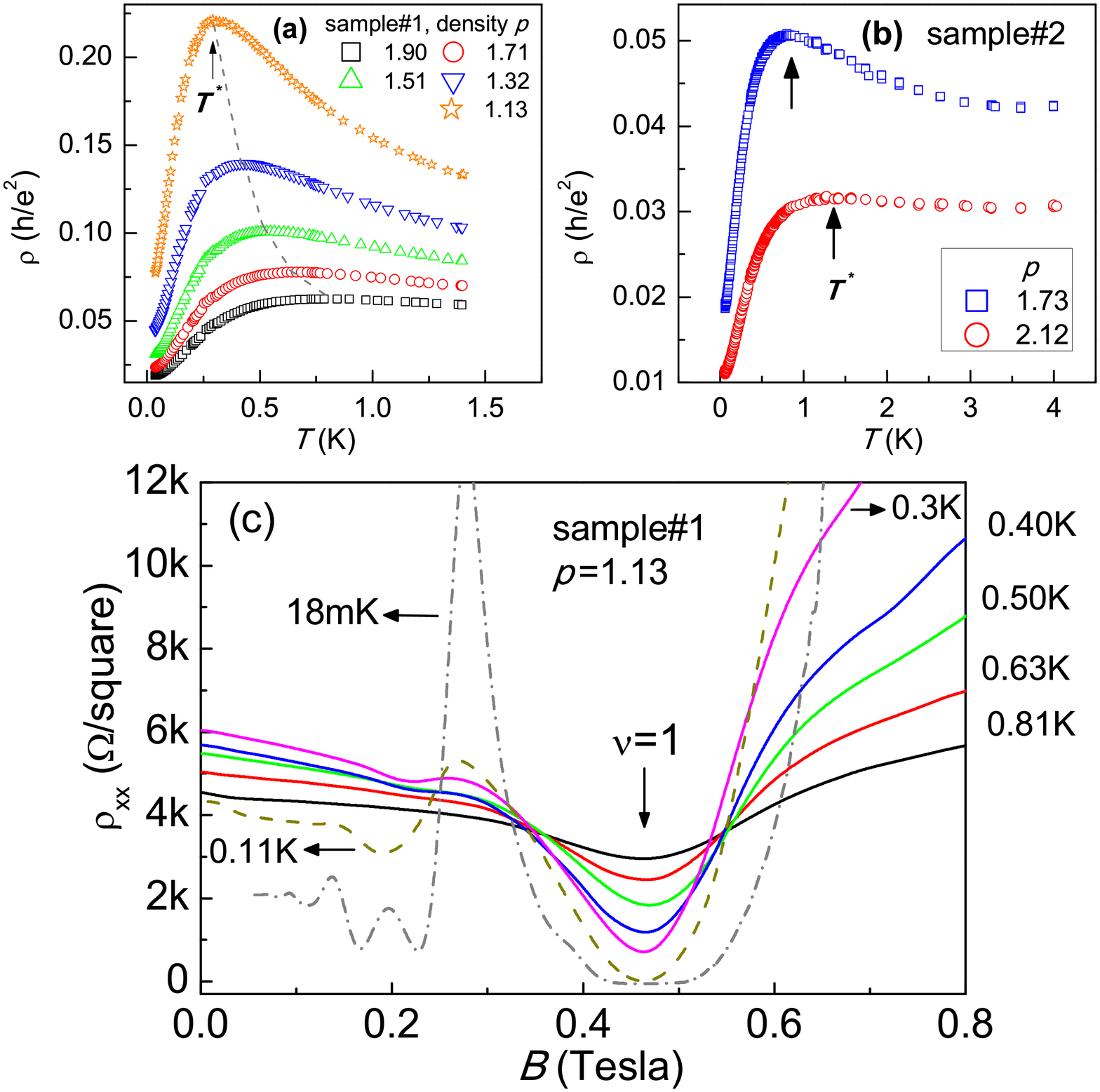,width=9cm}}
 \caption{(color online) (a)Non-monotonic $\rho(T)$ for 2DHS with density
 ($p$=1.90, 1.71, 1.51, 1.32, 1.13) in a 10nm wide
GaAs QW sample $\#$1. The cross-over temperature $T^*$ of the
non-monotonic $\rho(T)$ of different $p$ is connected by a dashed
line to guide the eye. (b) $\rho$ vs. $T$ for $p$=1.73, 2.12 in
sample $\#$2, showing that non-monotonic $\rho(T)$ exists even in
the highly conductive regime with $\rho$ as low as $\sim
0.01\times h/e^2$. (c) Resistivity $\rho_{xx}$ vs. perpendicular
magnetic field $B$ at different temperatures for $p$=1.13 in
sample$\#$1. For $T>T^*$, the $\rho_{xx}(B)$ curves still exhibit
SdH dip at $\nu$=1 whose position does not change with $T$.}
  \label{fig1}
\end{figure}

Before we go into the details of the density dependent
conductivity data and analysis, we use Fig.1 to establish some
basic transport and magneto-transport behavior of the low density
2DHS. Fig\ref{fig1}.a presents the temperature dependent
resistivity $\rho(T)$ at $B$=0 for several densities ($p$=1.90,
1.71, 1.51, 1.32, 1.13) in sample $\#$1 (the unit for $p$ is
10$^{10}/$cm$^{2}$ throughout this paper). For this sample, the
MIT happens around $p_c\approx$0.8 when the density is changed by
the back-gate voltage $V_g$ \cite{gaoHall}. It can be seen in
Fig.\ref{fig1}a that $\rho(T)$ changes from metallic
(d$\rho$/d$T>$0) to insulating-like (d$\rho$/d$T<$0) above a
characteristic temperature  $T^*$ which becomes larger when $p$
increases, consistent with previous findings in
literature\cite{Haneinprl,millsprl,gaoprl,Lilly}. Thanks to the
suppressed phonon scattering in our narrow QWs compared to wider
QWs or heterostructures\cite{millsprl}, here we are able to
directly observe such non-monotonic $\rho(T)$ into a much lower
resistivity regime ($\rho\sim$0.01$\times h/e^2$) than literature,
as shown in Fig.1b for $p$=2.12 in sample$\#$2.

The 2D hole density is determined by the positions of Shubnikov-de
Haas (SdH) oscillations as $p=\nu\times B_{\nu}\times e/h$, where
$\nu$ is the Landau filling factor and $B_{\nu}$ is the
perpendicular magnetic field at the corresponding $\nu$. Fig.1c
plots the longitudinal magneto-resistivity $\rho_{xx}(B)$ for
$p$=1.13 of sample $\#$1 at various temperatures
($T$=0.018-0.81K). Throughout this whole temperature range
covering both $T<T^*$ and $T>T^*$, SdH oscillation is well
established at $\nu$=1 and $B_{\nu=1}$ does not change with
temperature, indicating a constant $p$. In addition, we note that
that the $\nu$=1 SdH oscillation persists up to at least 0.81K, a
temperature comparable to $T_F$ (=1.0K using effective hole mass
$m^*$=0.3$m_e$\cite{mass}) or the cyclotron energy $\Delta_c$ at
$\nu$=1. This is surprising since SdH oscillation amplitude should
decay strongly above $k_BT\sim$0.1$\times\Delta_c$ according to
the Lifshitz-Kosevich formula\cite{SdHformula},
$\delta\rho_{xx}\propto\frac{2\pi^2k_BT/\Delta_c}{sinh(2\pi^2k_BT/\Delta_c)}$.
The fact that SdH is observed at $T>T^*$ points to the
non-classical nature of dilute 2DHS with large $r_s$ at these
semi-degenerate temperatures.

\begin{figure*}[btph]
\centerline{\psfig{file=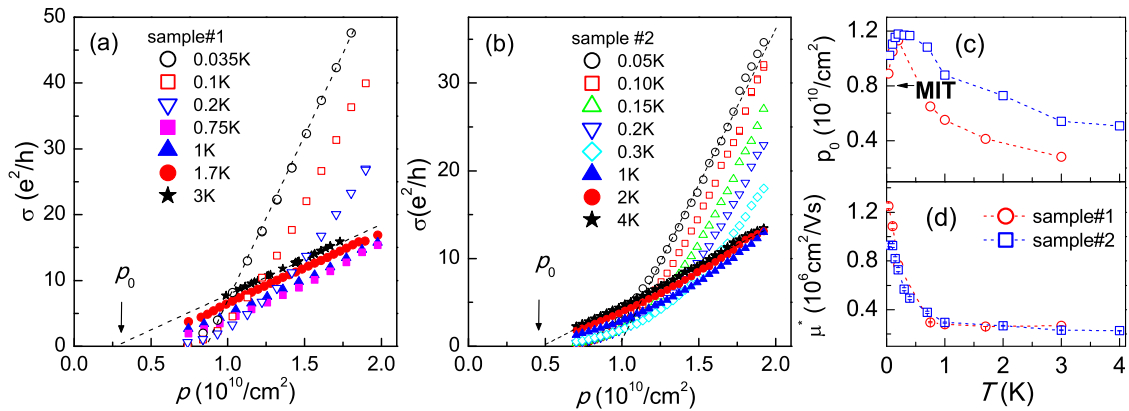,width=15cm}}
\caption{(color online) (a) The 2D hole conductivity $\sigma$
plotted as a function of density $p$ at different temperatures for
sample $\#$1 (a) and $\#$2 (b). A linear Drude-type function can
be used to approximate the conductivity as
$\sigma$=$e\mu^*(p-p_0)$, as denoted by the black dashed lines.
(c,d) Fitting parameters $p_0$ and $\mu^*$ plotted against
temperature.}\label{fig2}
\end{figure*}

While a decreasing $\rho$ with $T$ in the regime of $T>T^*$ is
expected in several models, either due to interaction/correlation
effects \cite{vlad,punnoose,spivakkivelson} or classical
scattering \cite{dassarma}, it has been difficult to identify the
exact mechanism\cite{Lai}. We studied the density dependence of
the conductivity $\sigma$ to gain more insights into the transport
mechanism of metallic 2DHS when the temperature coefficient
d$\rho$/d$T$ changes sign at $T^*$. Fig.2a and b present
$\sigma(p)$ for sample $\#$1 and $\#$2 from 0.035K ($T\ll T_F$) up
to 4K ($T>T_F$) over the density range 0.7$<p<$2. A few features
in $\sigma(p)$ are salient when the 2DHS crosses over from the low
$T$ degenerate (open symbols) to high $T$ semi-degenerate (solid
symbols) regime. First, similar to a previous report
\cite{hanein}, at low $T$'s, $\sigma(p)$ turns up sharply around
the MIT ($p=p_c$) and then follows a straight line with large
slope at $p>p_c$. As $T$ increases, the slope of the linear
dependence becomes smaller and at the same time the sharp upturn
at $p\sim p_c$ straightens. Eventually at high temperatures,
$\sigma(p)$ becomes a linear function over the whole range of $p$.
Yet the slope and intercept of the $\sigma(p)$ data at high $T$
are much smaller than the low $T$ curves. Linear $\sigma(p)$
dependence is expected in Drude model with the slope corresponding
to the mobility of carriers. Therefore, the dramatic slope change
in $\sigma(p)$ in our data suggests a dramatically enhanced
mobility of free carriers at low $T$. In the Drude model, the
finite intercept $p_0$ of $\sigma(p)$ would correspond to the
density of localized carriers, which appears to change with $T$ as
indicated by data in Fig.2a and b.

Previously, the low $T$ behavior of $\sigma(p)$ near the critical
regime ($p\sim p_c$) was analyzed in terms of the percolation
model as $\sigma=A(p-p_c)^\delta$ with $\delta\approx 4/3$ where
the MIT is driven by the percolation of itinerant carriers with
density $p-p_c$ through localized carriers with density $p_c$
\cite{Lai,Lillypercolation,Manfrapercolation,Tracypercolation}. A
linear relation ($\delta=1$) between $\sigma$ and $p-p_c$ at high
$\sigma$ would reconcile the percolation model with the Drude
formula when the overall conductivity is dominated by itinerant
carriers (i.e. when $\sigma\gg e^2/h$). In that case, the
coefficient $A$ in the percolation equation yields the effective
mobility $\mu^*$ of {\it itinerant} carriers. Here, we focus on
the the linear Drude part of $\sigma(p)$ in the high conductivity
limit ($\sigma\gg e^2/h$) to examine how the system evolves over a
broad range of $T$. This analysis not only gives a physically
meaningful parameter ($\mu^*$), but also works in the high
temperature (semi-degenerate) regime where the percolation fit is
not applicable. We fit the data in Fig.2a and b with
$\sigma>5e^2/h$ to $\sigma=e\mu^*(p-p_0)$ with $\mu^*$ and $p_0$
as the fitting parameters\cite{fitnote}. The fitted $\mu^*$ and
$p_0$ are plotted as a function of $T$ in Fig.2c and d for both
samples. First of all, reflecting the metallic transport and
rapidly increasing slope of $\sigma(p)$ at low $T$, $\mu^*$
exhibits a sharp upturn at $T$ lower than $\sim$ 0.5K, in contrast
to its nearly $T$-independent behavior at high $T$. On the other
hand, $p_0$ only shows minor drop at $T<$0.5K but decreases
greatly at high $T$'s (a factor of three/two for sample $\#$1/2).
These effects revealed through the {\it density-dependent}
conductivity analysis lead to an important insight on the
non-monotonic $\rho(T)$ or $\sigma(T)$ for metallic 2DHS with a
{\it fixed density}: while the metallic conduction in the
degenerate regime could be attributed to an enhanced mobility of
itinerant carriers, the increasing conductivity in the
semi-degenerate regime (i.e. $T>T^*$ in Fig.1) is a consequence of
decreased $p_0$, or the density of localized carriers, but not a
mobility effect. This can actually be inferred directly from the
raw data in Fig.2a: as $T$ is lowered from 3K to 0.75K, the
$\sigma(p)$ curves stay parallel to each other and shift towards
higher intercept ($p_0$). Within this analysis, one obtains the
following picture for the non-monotonic $\rho(T)$ peak around
$T^*$ in low density 2D systems: the resistivity drop at $T<T^*$
is due to reduced scattering but the high $T$ ($T>T^*$)
resistivity drop comes from a different mechanism where some
localized carriers become itinerant and contribute more and more
to the overall conductivity when the temperature is increased.

To explain the non-monotonic $\rho(T)$ of low density 2D carrier
systems, a few theories involving different mechanisms were
proposed \cite{dassarma,punnoose,spivakkivelson}. While it is
natural that these theories have focused on the effect of
temperature on the scattering, diffusion or viscosity of the 2D
carriers, our data supply two useful insights that are not
contained specifically in the existing theories. First, in the
high $T$ semi-degenerate regime, effective mobility $\mu^*$ of
itinerant carriers (essentially the slope of d$\sigma$/d$p$) is
roughly $T$-independent and much smaller than the degenerate
regime. This emphasizes a distinct transport property of the 2DHS
between $T<T^*$ and $T>T^*$: although the 2DHS can have same
resistivity value in the degenerate or semi-degenerate regime
(Fig.1a), carriers added to the system experience much stronger
scatterings/collisions at $T>T^*$ than $T<T^*$. Second, the
decreasing $\rho$ in the $T>T^*$ regime is likely tied with the
temperature dependence of $p_0$, the density of localized
carriers, instead of a simple scattering rate effect. These
features should be included in future theoretical considerations.

\begin{figure*}[btph]
\centerline{\psfig{file=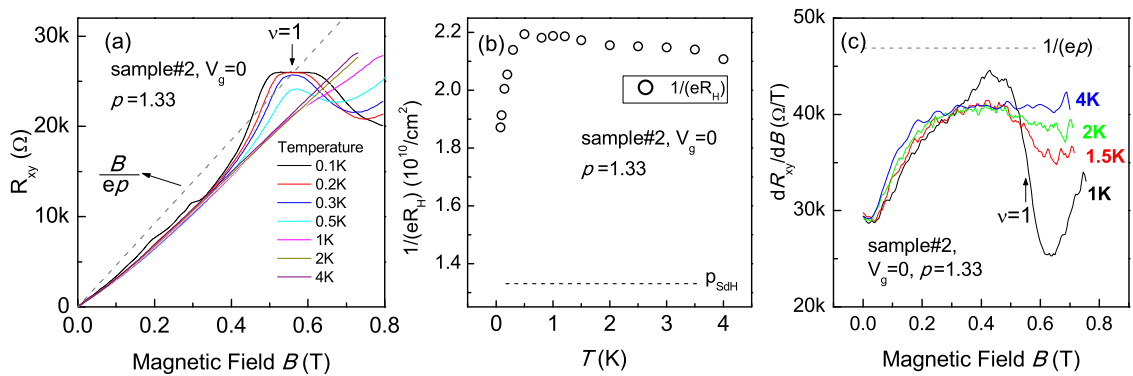,width=15cm}}
\caption{(color online) (a) Hall resistance $R_{xy}$ as a function
of perpendicular magnetic field $B$ for $p$=1.33 in sample $\#$2
at $T$=0.1-4K. (b) 1/($eR_H$) vs. $T$ for $p$=1.33 with $R_H$ as
the low field Hall coefficient. (c) Hall slope, d$R_{xy}$/d$B$ vs.
magnetic field at $T$=1, 1.5, 2 and 4K. At high $T$ when the
mixing of $\nu$=1 QH effect weakens, the Hall slope shows clear
enhancement as $B$ increases.}\label{fig3}
\end{figure*}
We studied the transverse magneto-resistance or Hall resistance
$R_{xy}$ in perpendicular magnetic field to obtain further
information on the nature of the two species of carriers inferred
from $\sigma(p)$ data. Fig.3a shows $R_{xy}$ vs. $B$ for $p$=1.33
of sample$\#$2 over a broad temperature range ($T$=0.1-4K). The
data have been symmetrized using both positive and negative field
measurements to remove the slight mixing from longitudinal
resistance. A grey dashed line is included to show the classical
linear Hall resistance $B/(ep)$ according to the hole density $p$.
It is curious to see that the experimentally measured $R_{xy}$ is
always smaller than $B/(ep)$, except in the fully developed
$\nu$=1 QH state. This significant difference between measured
$R_{xy}$ and $B/(ep)$ is quantified in Fig.3b and Fig.3c. Fig.3b
shows that 1/$eR_H$ is significantly (40-60$\%$) higher than $p$.
Here, $R_H$ is the Hall coefficient obtained by fitting the slope
of $R_{xy}(B)$ at low field ($|B|<$500Gauss). It is tempting to
relate the temperature dependence of 1/$eR_H$ in Fig.3b to a $T$
dependent carrier density effect similar to what we infer from
$\sigma(p)$ data. However, two caveats are worth to point out.
First, the increase of 1/($eR_H$) at $T<$1K reproduces previous
results on the $T$-dependent $R_H$ in similar 2DHS\cite{gaoHall}
whose origin is not fully understood since multiple mechanisms can
lead to temperature dependent corrections to $R_H$
\cite{zala,dassarmahall,yasin}. The second caution or puzzle one
needs to consider is the significant difference between 1/$eR_H$
and $p$: it is as large as 40$\%$ even at the lowest temperature
studied (80mK). Fig.3c presents the $B$-dependence of Hall slope
d$R_{xy}$/d$B$, up to 0.8T. It shows that although d$R_{xy}$/d$B$
exhibits some increase in $B$, it is still lower than 1/($ep$)
which is anticipated from density. In conventional two-band
transport model, the low field Hall slope d$R_{xy}$/d$B$ is
related to both the density and mobility of the two carrier
species as 1/($e\times$d$R_{xy}$/d$B$) = $(p_1\mu_1 + p_2\mu_2)^2
/(p_1\mu_1^2 + p_2\mu_2^2)$, while 1/($e\times$d$R_{xy}$/d$B$) at
high fields ($\mu B\gg$1) equals to the total carrier density $p_1
+ p_2$. Thus the standard two band model always expects a smaller
Hall slope at high field. We see that the {\it opposite} trend is
exhibited in our 2DHS as d$R_{xy}$/d$B$ becomes larger at higher
$B$ in Fig.3c. We speculate that the increase in Hall slope at
high $B$ is caused by the localization or Wigner crystallization
of carriers\cite{gabor}. Because the two carrier species we
consider can have both temperature and magnetic field dependent
density and mobility, we do not have a reliable model to fit
$R_{xy}(B)$ data to compare with the zero field conductivity
analysis in Fig.2. One possible implication of the small Hall
slope in our experiments is that there exists carriers which
contribute to current but not Hall voltage. Obviously, further
study is required to understand the anomalous Hall slope and the
nature of the two carrier species in 2DHS with large $r_s$.

It is worthwhile to point out that our results may in fact be
compatible with several theories which emphasize the coexistence
of a conducting metallic phase and an insulating localized phase
near the MIT \cite{wc,spivakkivelson,dassarma}. In the
micro-emulsion scenario of the 2D MIT, the 2D metallic phase
constitutes of mobile Fermi liquids percolating through bubbles of
Wigner crystals which have much lower conductivity
\cite{spivakkivelson}. The original micro-emulsion model suggested
that the 1/$T$-dependent viscosity of correlated electron fluid as
the explanation for the decreasing $\rho$ in the high $T$ regime
of $T\sim T_F$. Our $\sigma(p)$ data suggest that the continuous
melting of Wigner crystal is perhaps more important in the
experimentally accessed temperature range here, since only $p_0$
has strong temperature dependence at $T>T^*$. Thus theoretical
calculations on the density dependent conductivity and the Hall
effect of micro-emulsion would be desirable to compare further
with experiment. In other classical percolation models of the 2D
MIT \cite{Lillypercolation,Manfrapercolation,Tracypercolation},
the nature and the high temperature fate of the localized carriers
have not been addressed theoretically so far. In those theories,
more detailed calculations need be done to see if thermal
activation of localized carriers can produce the effects reported
here.

In summary, we have studied the density dependent conductivity of
a 2DHS in GaAs QW as $T$ is raised from the low $T$ degenerate
($T\ll T_F$) to high $T$ semi-degenerate ($T>T_F$) regime. In both
regimes, the system's conductivity can be described by a
Drude-like formula $\sigma(p)\approx e\mu^*(p-p_0)$ in the high
conductivity limit. The temperature dependence of $\sigma(p)$
reveals that the metallic transport at $T<T_F$ is associated with
the dramatically enhanced $\mu^*$ at low $T$, while the system's
resistivity decrease at $T\sim T_F$ is likely a result of some
localized carriers becoming conducting. However, the temperature
and magnetic field dependence of Hall resistance requires further
understanding.

X.P.A.G. is indebted to G.S. Boebinger, A.P. Mills and A.P.
Ramirez for useful suggestions at the early stage of the work, and
thanks E. Abrahams, S. Das Sarma, V. Dobrosavljevi$\acute{c}$, S.
Kivelson, and B. Spivak for discussions. This work was supported
by NSF (Grant No. DMR-0906415).


\begin{references}
\bibitem{Lee&rama}P. A. Lee and T. V. Ramakrishnan, Rev. Mod. Phys. {\bf 57}, 287 (1985).
\bibitem{scaling79}E. Abrahams, P. W. Anderson, D. C. Licciardello and T. V. Ramakrishnan, Phys. Rev. Lett. \textbf{42}, 673 (1979).
\bibitem{mitreview}B. Spivak, S. V. Kravchenko, S.A. Kivelson and X. P.A. Gao, Rev.
Mod. Phys. \textbf{82}, 1743 (2010);  E. Abrahams, S. V.
Kravchenko, and M. P. Sarachik, Rev. Mod. Phys. \textbf{73}, 251
(2001).
\bibitem{hexie}S. He and X. C. Xie, Phys. Rev. Lett. {\bf 80}, 3324 (1998).
\bibitem{philips}P. Phillips,Y. Wan, I. Martin, S. Knysh, and D. Dalidovich, Nature {\bf 395}, 253 (1998).
\bibitem{wc}S. Chakravarty,S. Kivelson,C. Nayak,K. Voelker, Philos. Mag. B {\bf 79}, 859 (1999).
\bibitem{vlad}A.Camjayi, K. Haule, V. Dobrosavljevi$\acute{c}$ and G. Kotliar, Nat. Phys. {\bf 4}, 932 (2008); M.C. O. Aguiar, E. Miranda, V. Dobrosavljevi$\acute{c}$, E. Abrahams and G. Kotliar, Europhys. Lett. {\bf
67}, 226 (2004).
\bibitem{punnoose}A. Punnoose and A. Finkel'stein, (a) Science {\bf 310}, 289
(2005); (b) Phys. Rev. Lett. {\bf 88}, 016802 (2001).
\bibitem{spivakkivelson} B. Spivak and S.A. Kivelson, Annals of Physics {\bf 321},
2071 (2006); Phys. Rev. B  {\bf 70}, 155114 (2004); B. Spivak,
Phys. Rev. B {\bf 67}, 125205 (2003).
\bibitem{altshuler}B. L. Altshuler, D. L. Maslov, and V. M. Pudalov, Physica E {\bf 9}, 209 (2001).
\bibitem{zna}G. Zala, B. N. Narozhny, and I. L. Aleiner, Phys. Rev. B  {\bf64}, 214204 (2001).
\bibitem{dassarma}S. Das Sarma and E.H. Hwang, Solid State Comm. {\bf 135}, 579
(2005); Phys. Rev. B {\bf 68}, 195315 (2003); Phys. Rev. B {\bf
61}, R7838 (2000); Phys. Rev. Lett. {\bf 83}, 164 (1999).
\bibitem{skvprb99}S. V. Kravchenko, D. Simonian, K. Mertes, M. P. Sarachik, and T. M. Klapwijk, Phys. Rev. B {\bf59}, R12740 (1999).
\bibitem{Lai}K. Lai, W. Pan, D.C. Tsui, S. Lyon, M. M$\ddot{u}$hlberger, and F. Sch$\ddot{a}$ffler, Phys. Rev. B {\bf 75},
033314 (2007).
\bibitem{Haneinprl} Y. Hanein, U. Meirav, D. Shahar, C.C. Li, D.C. Tsui, and H. Shtrikman, Phys. Rev. Lett. \textbf{80}, 1288 (1998).
\bibitem{millsprl} A. P. Mills, Jr., A. P. Ramirez, L. N. Pfeiffer, and K. W. West, Phys. Rev. Lett. \textbf{83}, 2805 (1999).
\bibitem{gaoprl}X. P.A. Gao, A. P. Mills, Jr., A. P. Ramirez, L. N. Pfeiffer, and K. W. West, Phys. Rev. Lett. (a) \textbf{88}, 166803 (2002); (b) \textbf{89}, 016801 (2002).
\bibitem{Lilly} M. P. Lilly, J. L. Reno, J.A. Simmons, I.B. Spielman, J.P. Eisenstein, L.N. Pfeiffer, K.W. West, E.H. Hwang, and S. Das Sarma, Phys. Rev. Lett. \textbf{90}, 056806 (2003).
\bibitem{gaophonon}X. P.A. Gao, G.S. Boebinger, A. P. Mills, Jr., A. P. Ramirez, L. N. Pfeiffer, and K. W. West, Phys. Rev. Lett. {\bf 94}, 086402 (2005).
\bibitem{novikov}D. S. Novikov Phys. Rev. B {\bf 79}, 235304 (2009).
\bibitem{mass}Calculation of $r_s$ depends on the effective hole mass $m^*$. A frequently
cited effective hole mass is $m^*$=0.35$m_e$ (H. L. Stormer and W.
T. Tsang, Appl. Phys. Lett. {\bf 36}, 685 (1980)). Recent
cyclotron resonance experiments on a 10nm wide GaAs QW with
somewhat higher density than our sample give $m^*=0.19m_e$ (W.
Pan, K. Lai, S.P. Bayrakci, N.P. Ong, D.C. Tsui, L.N. Pfeiffer,
and K.W. West, Appl. Phys. Lett. {\bf 83}, 3519 (2003)). By
fitting the $T$-dependent SdH amplitudes at low field to the
Lifshitz-Kosevich formula, we estimate $m^*\approx$0.3$m_e$ for
our sample.
\bibitem{gao06}X. P.A. Gao, G.S. Boebinger, A. P. Mills, Jr., A. P. Ramirez, L. N. Pfeiffer, and K. W. West, Phys. Rev. B {\bf 73}, 241315(R) (2006).
\bibitem{gaoHall}X. P.A. Gao, G.S. Boebinger, A. P. Mills, Jr., A. P. Ramirez, L. N. Pfeiffer, and K. W. West, Phys. Rev. Lett. {\bf 93}, 256402 (2004).
\bibitem{SdHformula}A. Isihara and L. Smr$\check{c}$ka, J. Phys. C {\bf 19}, 6777 (1986).
\bibitem{hanein}Y. Hanein, D. Shahar, J. Yoon, C.C. Li, D.C. Tsui, and H. Shtrikman, Phys. Rev. B {\bf 58}, R7520 (1998).
\bibitem{Lillypercolation}S. Das Sarma,M.P. Lilly, E.H. Hwang, L.N. Pfeiffer, K.W. West, and J.L. Reno, Phys. Rev. Lett. {\bf 94}, 136401 (2005).
\bibitem{Manfrapercolation}M. J. Manfra, E.H. Hwang, S. Das Sarma, L.N. Pfeiffer, K.W. West, and A.M. Sergent, Phys. Rev. Lett. {\bf 99}, 236402 (2007).
\bibitem{Tracypercolation}L. A. Tracy, E.H. Hwang, K. Eng, G.A. Ten Eyck, E.P. Nordberg, K. Childs, M.S. Carroll, M.P. Lilly, and S. Das Sarma, Phys. Rev. B {\bf 79},
235307 (2009).
\bibitem{fitnote}The exact choice of threshold conductivity (5$e^2/h$ in our fitting) in the linear fit of $\sigma(p)$ is somewhat arbitrary and does not affect the behavior of fitted
$\mu^*$ and $p_0$ vs. $T$ in a fundamental way.
\bibitem{zala}G. Zala, B.N. Narozhny, and I.L. Aleiner, Phys. Rev.
B {\bf 64}, 201201(R) (2001).
\bibitem{dassarmahall}S. Das Sarma and E.H. Hwang, Phys. Rev.
Lett. {\bf 95}, 016401 (2005).
\bibitem{yasin}C. E. Yasin, T. L. Sobey, A. P. Micolich, W. R. Clarke, A. R. Hamilton, M. Y.
Simmons, L. N. Pfeiffer, K. W. West, E. H. Linfield, M. Pepper,
and D. A. Ritchie Phys. Rev. B {\bf 72}, 241310 (2005).
\bibitem{gabor}G. A. Cs$\acute{a}$thy, Hwayong Noh, D. C. Tsui, L. N. Pfeiffer,
and K. W. West, Phys. Rev. Lett. {\bf 94}, 226802 (2005).

\end{references}
\end{document}